\documentclass[english,twocolumn,pra,aps,groupedaddress,showpacs]{revtex4-1}
\usepackage{lmodern}
\usepackage[T1]{fontenc}
\usepackage[latin9]{inputenc}
\usepackage{color}
\usepackage{babel}
\usepackage{float}
\usepackage{amsmath}
\usepackage{amssymb}
\usepackage{graphicx}
\usepackage{esint}
\usepackage[unicode=true,
 bookmarks=false,bookmarksnumbered=false,bookmarksopen=false,
 breaklinks=false,pdfborder={0 0 1},backref=false,colorlinks=true]
 {hyperref}
\hypersetup{pdftitle={Resummation of infrared divergencies in the theory of atomic Bose
 gases},
 pdfauthor={H.T.C. Stoof and J.J.R.M. van Heugten},
 citecolor=blue, linkcolor=blue, urlcolor=blue}

\makeatletter
 
 \@ifundefined{textcolor}{}
 {%
   \definecolor{BLACK}{gray}{0}
   \definecolor{WHITE}{gray}{1}
   \definecolor{RED}{rgb}{1,0,0}
   \definecolor{GREEN}{rgb}{0,1,0}
   \definecolor{BLUE}{rgb}{0,0,1}
   \definecolor{CYAN}{cmyk}{1,0,0,0}
   \definecolor{MAGENTA}{cmyk}{0,1,0,0}
   \definecolor{YELLOW}{cmyk}{0,0,1,0}
 }

\usepackage{bbold}
\renewcommand{\fnum@figure}{FIG.~\thefigure}

\makeatother

\begin{document}

\title{Resummation of infrared divergencies in the theory of atomic Bose gases}

\author{H.T.C. Stoof and J.J.R.M. van Heugten}

\affiliation{Institute for Theoretical Physics, Utrecht University, Leuvenlaan
4, 3584 CE Utrecht, The Netherlands}
\begin{abstract}
We present a general strong-coupling approach for the description
of an atomic Bose gas beyond the Bogoliubov approximation, when infrared divergences start to occur that need to be resummed exactly. We consider the determination of several important physical properties of the Bose gas, namely the chemical potential, the contact, the speed of sound, the condensate density, the effective interatomic interaction and the
three-body recombination rate. It is shown how the approach can be systematically
improved with renormalization-group methods and how it reduces to
the Bogoliubov theory in the weak-coupling limit.
\end{abstract}

\pacs{67.85.-d, 67.10.Ba, 03.75.-b}

\maketitle
The main challenge of statistical physics is to describe the many-body
properties of a system given the underlying few-body physics. Cold
atomic gases provide a versatile experimental testbed for these theoretical
descriptions by allowing the investigation of the crossover of many-body
systems from weak to strong two-body interactions, using magnetic-field-tunable
Feshbach resonances \citep{Chin2010Feshbach,Bloch2008Manybody,Stoof2009Ultracold}.
In particular, the universal nature of fermionic many-body systems with
resonant two-body interactions has been successfully studied experimentally
and theoretically \citep{Zwer2011BECBCS,Bloch2008Manybody}. The most
remarkable property of such resonant systems, which have an infinite
scattering length and are therefore said to be at unitarity, is that
at zero temperature there is no other length scale than the average
interatomic distance that is set by the particle density $n$. As
a result all thermodynamic quantities, when appropriately scaled,
can be expressed in terms of a set of universal numbers. For the case
of the Fermi gas at unitarity, one of the most crucial quantities
is the chemical potential
\begin{equation}
\mu=(1+\beta)\epsilon_{F},\label{eq:Chemical_potential}
\end{equation}
which is given by an universal constant times the Fermi energy $\epsilon_{F}=\hbar^{2}k_{F}^{2}/2m$,
where $k_{F}=(6\pi^{2}n/2s+1)^{1/3}$ is the Fermi momentum and $s=1/2$
due to the hyperfine degrees of freedom. The universal constant $\beta$
can be interpreted as describing the deviation from the ideal gas
result due to interactions and was found to be $\beta\simeq-0.63$
experimentally as well as theoretically \citep{Zwer2011BECBCS,Ku2012Revealing,Zurn2012Precise}.

Recently there has been increasing experimental interest in the strongly
interacting Bose gas \citep{Papp2008Bragg,Pollack2009Extreme,Navon2011Dynamics,Wild2012Measurements, Rem2013Lifetime,Zoran2013}.
It is expected on dimensional grounds that the Bose gas at unitarity,
if stable, has similar universal properties as that of the unitary
Fermi gas. For instance Eq.~(\ref{eq:Chemical_potential}) is expected
to hold also but with $s=0$ and a different value of $\beta$ due
to the different statistics of the atoms. In contrast to the unitary
Fermi gas, the realization of the unitary Bose gas is complicated
by an increased loss of atoms as a consequence of a strong increase
in the rate of inelastic three-body recombination processes caused
by the absence of the Pauli principle and the existence of Efimov
trimers. These three-body processes result in the formation of molecules,
which shows that the actual ground state of these gases is a Bose-Einstein
condensate of molecules. Nevertheless, it may still be experimentally
possible to create the meta-stable state of a Bose-Einstein condensate
of atoms at large scattering lengths for a sufficiently long time \citep{Zoran2013}.
We have little to say about this important problem in this paper, and assume from now on that such a meta-stable state can indeed be realized in the laboratorium.

On the theoretical side, the description of the unitary Bose gas has
been challenging and recent theoretical results strongly vary \citep{Cowell2002Cold,Yin2008Phase,Song2009Ground,Lee2010Universality, Diederix2011Ground,Li2012Bose,Borzov2012Threedimensional,Adhikari2008Nonlinear,
Cooper2010Nonperturbative,Zhou2013,Fei2013}.
The main difficulty with constructing a theory of unitary Bose gases
comes from the fact that there is no small parameter in the theory.
Variational studies circumvent this by finding the minimum of the
thermodynamic potential. However, since we are interested in the meta-stable
state, care should be taken to project out the true many-body groundstate.
In addition, diagrammatic approaches beyond the Bogoliubov theory are known to be plagued by logarithmic infrared divergences, as was
first noted by Gavoret and Nozi\'eres \citep{Gavoret1964Structure}.

Motivated by these ongoing efforts to study atomic Bose gases with strong interaction effects, the main objective of this paper is to present a general strong-coupling
approach to an interacting Bose gas that can be improved systematically, for instance by renormalization-group methods but also by other non-pertubative methods such as the large-$N$ expansion. Our approach is by construction free of the troublesome infrared
divergences by exactly incorporating the phase fluctuations of the
Bose-Einstein condensate, which are known to dominate the long-wavelength
behavior of the system \citep{Gavoret1964Structure,Popov1971Application,Popov1972Hydrodynamic}.
More precisely, the theory is first renormalized by all other fluctuations
using for instance the renormalization group. Then using this improved
theory we next include the effects of the phase fluctuations of the
Bose-Einstein condensate, which is reminiscent of bosonization for
fermions. That the phase fluctuations are exactly incorporated will
be confirmed by reproducing the exact form of the single-particle
propagator in the long-wavelength limit as derived by Nepomnyashchii
and Nepomnyashchii \citep{Nepomnyashchii1975Contribution,Nepomnyashchii1978Infrared}.

The outline of the paper is as follows. In section \ref{sec:Beyond-Bogoliubov}
we give a brief overview of Bogoliubov theory and
discuss the difficulties in going beyond this theory, such
as the appearance of the above-mentioned infrared divergences. Subsequently,
in section \ref{sec:Renormalized-Bosonization}, we present our strong-coupling
approach which circumvents these difficulties by incorporating the
phase fluctuations of the Bose-Einstein condensate exactly. In particular, the theoretical framework is discussed in subsection \ref{sub:Theory}. In section \ref{sub:Results} it is first discussed how the Bogoliubov theory is reproduced within this general framework when taking the weak-coupling limit. Next, we discuss as a proof of principle also a first non-trivial approximation that goes beyond the Bogoliubov theory and allows us to obtain finite results for several properties of the Bose gas as a function of the coupling constant, i.e., the scattering length. Finally we conclude our paper in section \ref{concl} and discuss various avenues for further improvement.

\section{Bogoliubov theory and beyond\label{sec:Beyond-Bogoliubov}}

In this section we illustrate the difficulties in constructing a theory
of the strongly interacting Bose gas, which will be of use when presenting our approach in section \ref{sec:Renormalized-Bosonization}.
We first briefly review in section \ref{sub:Bose_gas} Bogoliubov theory as a benchmark for our theory. Next, we recognize that the correct
low-energy behavior must be exactly incorporated into the theory. To do so
requires going beyond Bogoliubov theory at which point we encounter
the above-mentioned logarithmic infrared divergences, which are discussed in section~\ref{sub:Difficulties_Bog}.

\subsection{Atomic Bose gas\label{sub:Bose_gas}}

Here we briefly summarize some of the results of Bogoliubov theory,
including the first quantum corrections, as a benchmark for our theory.
In general, the Bose gas in cold atom experiments is well described
by the Euclidean action $S\left[\phi^{*},\phi\right]=\int\mathrm{d}\tau\mathrm{d}\mathbf{x}\,{L}(\mathbf{x},\tau)$
with a point interaction, where the lagrangian density is
\begin{align}
{L}(\mathbf{x},\tau) & =\phi^{*}(\mathbf{x},\tau)\left[\hbar\partial_{\tau}-\frac{\hbar^{2}\nabla^{2}}{2m}-\mu\right]\phi(\mathbf{x},\tau)\nonumber \\
 & +\frac{1}{2}T^{\mathrm{2B}}\left|\phi(\mathbf{x},\tau)\right|^{4}.\label{eq:Action_Bose_Gas-1}
\end{align}
Here $\phi$ is the atomic field, $\mu$ is the chemical potential, $T^{\mathrm{2B}}=4\pi a(B)\hbar^{2}/m$
is the exact two-body T(ransition) matrix at zero energy and momentum, $a(B)$ is the magnetic-field-tunable scattering length, and $m$ is the mass of the atoms.

In mean-field theory, which amounts to expanding the field in terms
of the condensate and neglecting the fluctuations around it, the time-independent
equation for the atomic condensate is
\begin{equation}
\mu=nT^{\mathrm{2B}}=\frac{\hbar^{2}}{ma^2}(4\pi na^3),\label{eq:MF_chemPot1}
\end{equation}
where it was used that at this level of approximation the condensate
density $n_{c}$ is equal to the total density $n$. The first quantum
correction to the above result was calculated by Lee-Huang-Yang (LHY)
using the Bogoliubov theory that also incorporates the gaussian fluctuations
around the mean-field solution, and results in \citep{Lee1957Eigenvalues}
\begin{equation}
\mu=\frac{\hbar^{2}}{ma^{2}}(4\pi na^{3})\left(1+\frac{16}{3\pi}\sqrt{4\pi na^{3}}\right).\label{eq:LHY_Chempot}
\end{equation}
The condensate density to this order is given by
\begin{equation}
n_{c}=n\left(1-\frac{4}{3\pi}\sqrt{4\pi na^{3}}\right).\label{eq:depletion_condensate}
\end{equation}
This shows the depletion from the condensate due to the interaction.
Higher-order corrections to the chemical potential have been determined
\citep{Braaten2002Dilute}, however, these depend also on three-body
physics and will not be discussed in detail here.

Another important quantity of the atomic Bose gas is called the contact
$C$ \citep{Tan2008Energetics,Tan2008Large,Braaten2008Exact,Combescot2009Particle, Schakel2010Tan,Werner2012General}.
It is determined by the short-wavelength behavior of the single-particle
distribution function, namely $n(\mathbf{k})\simeq C/\mathbf{k}^{4}$.
In Bogoliubov theory the contact is given by \citep{Schakel2010Tan}
\begin{equation}
C=(4\pi na)^{2}\left(1+\frac{48}{3\pi}\sqrt{4\pi na^{3}}\right),\label{eq:Bog_contact}
\end{equation}
where also the first quantum correction is shown, consistent with
the Lee-Huang-Yang correction of the chemical potential.

Clearly all the above quantities diverge in the unitarity limit $a\rightarrow\infty$,
which is not surprising since they are expansions in terms of the
small parameter $\sqrt{4\pi na^{3}}$. This is a consequence of the fact
that in Bogoliubov theory no many-body corrections on the scattering
length have been taken into account, such that the effective interaction
cannot become finite at unitarity. Therefore, to be able to describe
the Bose gas in the strongly-interacting limit ($na^{3}\gg1$) the
action $S[\phi^{*},\phi]$ needs to be properly renormalized as we discuss in much more detail later on.

\subsection{Difficulties beyond Bogoliubov theory\label{sub:Difficulties_Bog}}

In Bogoliubov theory it thus appears that we cannot reach the
strongly interacting regime. Therefore, we now want to go beyond Bogoliubov theory.
To correctly describe the low-energy behavior of the Bose-Einstein
condensate it appears natural to use Bogoliubov theory to describe
the excitations above the condensate. However, it proves difficult
to renormalize the action of the atomic Bose gas of Eq.~(\ref{eq:Action_Bose_Gas-1})
using the Bogoliubov propagator. Here we discuss some of the difficulties
we encounter when trying to renormalize the action after using the
Bogoliubov substitution. Again we expand the field around the condensate
density $n_{c}$, i.e.,
\begin{equation}
\phi(\mathbf{x},\tau)=\sqrt{n_{c}}+\phi'(\mathbf{x},\tau),\label{eq:Bog_substitution}
\end{equation}
and we obtain the mean-field equation
\[
\mu=n_{c}T^{\mathrm{2B}}.
\]
In Bogoliubov theory only terms quadratic in the fluctuations are
taken into account in the lagrangian, giving for the fluctuations
the action
\begin{equation}
\frac{1}{2}\sum_{\mathbf{k},n}{\mathbf{\Phi}'}^{\dagger}(\mathbf{k},\omega_{n})\left[-\hbar\mathbf{G}^{-1}(\mathbf{k},\omega_{n})\right]{\mathbf{\Phi}'}(\mathbf{k},\omega_{n}),\label{eq:Bog_Gaussian_action}
\end{equation}
with $\mathbf{\Phi}'(\mathbf{k},\omega_{n})=\left[\phi'(\mathbf{k},\omega_{n}),{\phi'}^{*}(-\mathbf{k},-\omega_{n})\right]^{T}$.
The components of the $2\times2$ (Nambu space) inverse Green's function
$\mathbf{G}^{-1}$ are
\begin{align}
 & -\hbar G_{11}^{-1}(\mathbf{k},\omega_{n})=-i\hbar\omega_{n}+\epsilon_{\mathbf{k}}-\mu+\hbar\Sigma_{11},\nonumber \\
 & \qquad\hbar\Sigma_{11}=2n_{c}T^{\mathrm{2B}},\quad\hbar\Sigma_{12}=n_{c}T^{\mathrm{2B}},\label{eq:Bog_SelfEn}
\end{align}
with the properties $G_{11}^{-1}(\mathbf{k},\omega_{n})=G_{22}^{-1}(-\mathbf{k},-\omega_{n})$
and also $-\hbar G_{12}^{-1}(\mathbf{k},\omega_{n})=-\hbar G_{21}^{-1}(\mathbf{k},\omega_{n})=\hbar\Sigma_{12}$. Moreover, $\omega_n$ are the bosonic Matsubara frequencies and $\epsilon_\mathbf{k} = \hbar^2 \mathbf{k}^2/2m$ is the free atomic dispersion.
Taking the inverse of the matrix in Eq.~(\ref{eq:Bog_Gaussian_action})
we obtain the $2\times2$ Bogoliubov Green's function, whose components
are
\begin{eqnarray}
-\hbar^{-1}G_{11}^{}(\mathbf{k},\omega_{n}) & = & \frac{i\hbar\omega_{n}+\epsilon_{\mathbf{k}}+n_{c}T^{\mathrm{2B}}}{-(i\hbar\omega_{n})^{2}+(\hbar\omega_{\mathbf{k}})^{2}},\label{eq:Bog_propagators}\\
-\hbar^{-1}G_{12}^{}(\mathbf{k},\omega_{n}) & = & \frac{-n_{c}T^{\mathrm{2B}}}{-(i\hbar\omega_{n})^{2}+(\hbar\omega_{\mathbf{k}})^{2}},\nonumber
\end{eqnarray}
where the mean-field equation was used to eliminate the chemical potential
and we defined the dispersion $\hbar\omega_{\mathbf{k}}$ as
\begin{equation}
\hbar\omega_{\mathbf{k}}=\sqrt{\epsilon_{\mathbf{k}}\left(\epsilon_{\mathbf{k}}+2n_{c}T^{\mathrm{2B}}\right)}.
\end{equation}

To go beyond the Bogoliubov approximation, which as we have seen is
necessary to describe a strongly interacting Bose gas, we need to
compute the corrections to the propagator, or more precisely to the self-energy matrix $\hbar \mathbf{\Sigma}$. Doing so, the one-loop correction gives rise to an infrared logarithmic divergency in the normal and anomalous self-energy as a consequence of the linear mode in the normal
and anomalous propagators, as was previously noted in Refs. \citep{Gavoret1964Structure,Dupuis2011Infrared}.
This is easily shown by realizing that at low momenta and low frequencies
both the normal and anomalous propagator are of the relativistic form
$1/{K}^{2}$ with the four-vector ${K}=(i\hbar\omega_{n},\sqrt{2n_{c}T^{\mathrm{2B}}\epsilon_{\mathbf{k}}})$
and thus first-order corrections to the normal and anomalous self-energies
give rise to a logarithmically divergent quantity $\Delta\Sigma$ proportional to
\[
\int\mathrm{d}^{4}{K}'\frac{1}{{K}'^{2}({K}'-{K})^{2}}\propto\log\left[\frac{-(i\hbar\omega_{n})^{2}+2n_{c}T^{\mathrm{2B}}\epsilon_{\mathbf{k}}}{\Lambda^{2}}\right],
\]
where $\Lambda$ is some high-energy cut-off obeying ${K}^{2}\ll\Lambda^{2}$.
This logarithmic divergence makes it increasingly difficult to apply
a self-consistent diagrammatic renormalization procedures to find the effective interaction and self-energies of the atoms. Nevertheless, it was shown by Nepomnyashchii
and Nepomnyashchii that an important consequence of these divergencies
is that the exact anomalous self-energy vanishes for zero momentum
and energy, i.e., $\hbar\Sigma_{12}(\mathbf{0},0)=0$ \citep{Nepomnyashchii1975Contribution,Nepomnyashchii1978Infrared}.
This indicates another difficulty with the Bogoliubov substitution,
since it gives rise to a non-zero anomalous self-energy, as for example
in Eq.~(\ref{eq:Bog_SelfEn}).

In general, when encountering infrared divergencies we need to perform
a resummation of an infinite amount of diagrams. Indeed, a resummation
of the one-loop diagrams gives in the long-wavelength limit $\hbar\Sigma_{11}=\mu+\Delta\Sigma^{-1}+{O}(\omega,\epsilon_{\mathbf{k}})$
and $\hbar\Sigma_{12}=\Delta\Sigma^{-1}+{O}(\omega^{2},\epsilon_{\mathbf{k}})$
\citep{Nepomnyashchii1978Infrared}, where $\Delta\Sigma$ is again the
above logarithm. Thus after resummation the anomalous self-energy
satisfies the exact relation $\hbar\Sigma_{12}(\mathbf{0},0)=0$.
Also, to obtain a consistent theory of the Bose gas it is necessary
to make sure that the theory has a gapless mode at each level of approximation
as a consequence of Goldstone's theorem. This statement is equivalent
to demanding that the self-energies satisfy the Hugenholtz-Pines relation
$\hbar\Sigma_{11}(\mathbf{0},0)-\hbar\Sigma_{12}(\mathbf{0},0)=\mu$
\citep{Hugenholtz1959GroundState}. The resummed self-energies indeed
satisfy this relation, quite simply as $ $$\hbar\Sigma_{11}(\mathbf{0},0)=\mu$.

Now, we may think that because we have obtained reasonable self-energies,
we are in a position to further investigate the effects of interactions.
This turns out to be no simple task, especially since the full self-energies
are quite involved. As an example, in order to re-obtain the sound
mode in the propagators in the long-wavelength limit it is already
necessary to deal with precise cancellations of the logarithms, as
was shown by Nepomnyashchii and Nepomnyashchii \citep{Nepomnyashchii1978Infrared}.

To summarize, after the Bogoliubov substitution we encounter difficulties
to go beyond the Bogoliubov approximation because of logarithmic infrared
divergencies. To perform self-consistent calculations of the effective
interaction and the normal and anomalous self-energies that always
satisfy the Hugenholtz-Pines relation and the requirement of a linear
mode in the single-particle Green\textquoteright{}s function quickly
becomes practically unfeasible. In the following we will isolate these
troublesome infrared divergences, which will be seen to originate
from the phase fluctuations of the Bose-Einstein condensate, and most importantly show
how to exactly incorporate these fluctuations in our approach.

\section{Renormalized Bosonization\label{sec:Renormalized-Bosonization}}

Here the general framework of our strong-coupling approach is presented.
Subsequently, after discussing the weak-coupling limit where the Bogoliubov theory is reproduced, we discuss a first non-trivial application of the general framework
to obtain several properties of the Bose gas as a function of scattering
length, such as the chemical potential, the contact, the speed of
sound, the condensate density and the effective interatomic interaction.
Lastly, we also discuss the unitarity-limited three-body recombination rate.

\subsection{Theory\label{sub:Theory}}

\begin{figure}[t]
\begin{centering}
\includegraphics[scale=0.7]{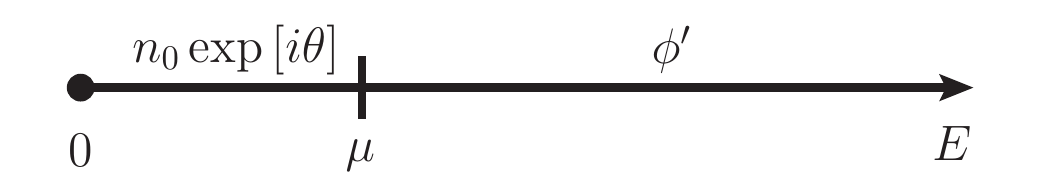}
\par\end{centering}

\caption{Schematic representation of the expansion of the field $\phi$ in
terms of the condensate and its phase fluctuations and the non-phase
fluctuations $\phi'$, c.f. Eq.~(\ref{eq:Expansion_of_field-1}).\label{fig:Expansion_wrt_scale}}
\end{figure}

In view of the problems discussed in the previous section, we now
show how to incorporate the phase fluctuations exactly and automatically
resum all infrared divergences in the theory. To describe the Bose-Einstein
condensed phase, we expand the field as
\begin{equation}
\phi(\mathbf{x},\tau)=\sqrt{n_{0}(\mathbf{x},\tau)}\exp\left[i\theta(\mathbf{x},\tau)\right]+\phi'(\mathbf{x},\tau),\label{eq:Expansion_of_field-1}
\end{equation}
where $n_{0}=\left\langle n_{0}(\mathbf{x},\tau)\right\rangle $ should
now be viewed as the quasicondensate density \citep{Stoof2009Ultracold}
and not as the density of atoms in the condensate $n_{c}$. The latter
will be related to $n_{0}$ by the large-distance behavior of the
fluctuations in the phase of the condensate $\theta(\mathbf{x},\tau)$ as we will see shortly.
Roughly speaking, the first term of the expansion describes the low-energy
modes of the field, as shown in Fig.~\ref{fig:Expansion_wrt_scale},
and includes the phase fluctuations. The fluctuations $\phi'(\mathbf{x},\tau)$
describe the high-energy modes and are defined such that they do not
contain phase fluctuations. The non-phase fluctuations $\phi'$ are
thus orthogonal to the first term in Eq.~(\ref{eq:Expansion_of_field-1}).
By inserting the expansion into Eq.~(\ref{eq:Action_Bose_Gas-1}),
the action $S\left[n_{0},\theta,\phi'^{*},\phi'\right]$ is obtained.

To proceed, we first show how to obtain the exact phase-fluctuation
propagator and the propagator of non-phase fluctuations from this
action. The latter will then be used to renormalize the theory using
the renormalization group, after which the exact contributions of
the phase fluctuations are re-introduced.

\subsubsection*{Propagator of the phase fluctuations}

The action for the phase fluctuations can be found by eliminating
the phase dependence of the part of the action involving $\phi'$
through the replacement
\[
\phi'(\mathbf{x},\tau)\rightarrow\exp\left[i\theta(\mathbf{x},\tau)\right]\phi''(\mathbf{x},\tau).
\]
This procedure of extracting the overall phase of the field $\phi$
is reminiscent of bosonization for fermions. The phase-fluctuation-dependent
part of the action $S\left[n_{0},\theta,\phi''^{*},\phi''\right]$
reduces to
\begin{align*}
 & \int\mathrm{d}\tau\mathrm{d}\mathbf{x}\left\{ \left[n_{0}(\mathbf{x},\tau)+\left|\phi''(\mathbf{x},\tau)\right|^{2}\right](i\hbar\partial_{\tau})\theta(\mathbf{x},\tau)\right.\\
 & \left.+\frac{\hbar^{2}}{2m}\left[n_{0}(\mathbf{x},\tau)+\left|\phi''(\mathbf{x},\tau)\right|^{2}\right](\nabla\theta(\mathbf{x},\tau))^{2}\right\} .
\end{align*}
As described above, it was used that $\phi''(\mathbf{x},\tau)$ and
$n_{0}(\mathbf{x},\tau)$ or $\theta(\mathbf{x},\tau)$ are
orthogonal to each other, i.e., the space-time integral over
their products vanish. Then by performing the path integral over the non-phase fluctuations $\phi''$, the phase-fluctuation-dependent part of the action in lowest order in the derivatives is
\begin{equation}
\int\mathrm{d}\tau\mathrm{d}\mathbf{x}\left\{ n(\mathbf{x},\tau)(i\hbar\partial_{\tau})\theta(\mathbf{x},\tau)
+\frac{\hbar^{2}n(\mathbf{x},\tau)}{2m}(\nabla\theta(\mathbf{x},\tau))^{2}\right\} .
\nonumber
\end{equation}
Here we introduced the total density
\begin{align*}
n(\mathbf{x},\tau) & =n_{0}(\mathbf{x},\tau)+\left\langle \phi''(\mathbf{x},\tau)\phi''^{*}(\mathbf{x},\tau)\right\rangle \\
 & =n_{0}(\mathbf{x},\tau)+\left\langle \phi'(\mathbf{x},\tau)\phi'^{*}(\mathbf{x},\tau)\right\rangle .
\end{align*}
Expanding the latter around its equilibrium value $n(\mathbf{x},\tau)=n+\delta n(\mathbf{x},\tau)$
the gaussian part of the action can be written in momentum space as
\begin{equation}
\frac{1}{2}\sum_{\mathbf{k},n}\left[\begin{array}{c}
\delta n(\mathbf{k},\omega_{n})\\
\theta(\mathbf{k},\omega_{n})
\end{array}\right]^{\dagger}\left(\begin{array}{cc}
\chi_{nn}(\mathbf{k}) & -\hbar\omega_{n}\\
\hbar\omega_{n} & 2n\epsilon_{\mathbf{k}}
\end{array}\right)\left[\begin{array}{c}
\delta n(\mathbf{k},\omega_{n})\\
\theta(\mathbf{k},\omega_{n})
\end{array}\right],\label{eq:ActionLowEnergy}
\end{equation}
where we introduced the exact density-density correlation function
$\chi_{nn}(\mathbf{k})$. The phase-fluctuation propagator is thus
found to be
\begin{eqnarray}
\left\langle \theta(\mathbf{k},\omega_{n})\theta^{*}(\mathbf{k},\omega_{n})\right\rangle  & = & \frac{\frac{1}{n}mc^{2}}{(\hbar\omega_{n})^{2}+2mc^{2}\epsilon_{\mathbf{k}}},\label{eq:Phase_fluc_prop-1}
\end{eqnarray}
where the speed of sound is $c=\sqrt{n\chi_{nn}(\mathbf{0})/m}$.
Note that we have obtained in this manner the exact phase-fluctuation propagator in the long-wavelength limit.

\subsubsection*{Propagator of the non-phase fluctuations}

When integrating out the non-phase fluctuations $\phi'$ the phase
of the condensate must be considered as non-fluctuating. Therefore,
the propagator of the non-phase fluctuations can be determined from
the action with a constant phase. Comparing the expansions of the
field in Eq.~(\ref{eq:Expansion_of_field-1}) with Eq.~(\ref{eq:Bog_substitution})
we see that the quadratic part of the action $S\left[n_{0},\theta,\phi'^{*},\phi'\right]$
with constant phase, for simplicity take $\theta=0$, is given by
the Bogoliubov action of Eq.~(\ref{eq:Bog_Gaussian_action}). The
usual Bogoliubov propagators, however, contain contributions of the
phase fluctuations, which can be identified by their proportionality
to $n_{0}$, since in Bogoliubov theory the phase fluctuations are described by $\sqrt{n_{0}}\exp\left[i\theta(\mathbf{x},\tau)\right]-\sqrt{n_{0}}\simeq i\sqrt{n_{0}}\theta(\mathbf{x},\tau)$.
We thus see that the contributions from the phase fluctuations are
\begin{eqnarray*}
\left\langle \phi'(\mathbf{x},\tau)\phi'^{*}(\mathbf{x},\tau)\right\rangle  & \propto & n_{0}\left\langle \theta(\mathbf{x},\tau)\theta(\mathbf{x},\tau)\right\rangle ,\\
\left\langle \phi'(\mathbf{x},\tau)\phi'^{*}(\mathbf{x},\tau)\right\rangle  & \propto & -n_{0}\left\langle \theta(\mathbf{x},\tau)\theta(\mathbf{x},\tau)\right\rangle .
\end{eqnarray*}
Thus we can remove the phase fluctuations from the Bogoliubov propagators
in Eq.~(\ref{eq:Bog_propagators}) by writing
\begin{align*}
\left\langle \phi'(\mathbf{k},\omega_{n})\phi'^{*}(\mathbf{k},\omega_{n})\right\rangle  & =\hbar\frac{i\hbar\omega_{n}+\epsilon_{\mathbf{k}}+n_{0}T^{\mathrm{2B}}}{(\hbar\omega_{n})^{2}+(\hbar\omega_{\mathbf{k}})^{2}}\\
 & -\hbar\frac{n_{0}T^{\mathrm{2B}}}{(\hbar\omega_{n})^{2}+(\hbar\omega_{\mathbf{k}})^{2}},\\
\left\langle \phi'(\mathbf{k},\omega_{n})\phi'(\mathbf{k},\omega_{n})\right\rangle  & =\hbar\frac{-n_{0}T^{\mathrm{2B}}}{(\hbar\omega_{n})^{2}+(\hbar\omega_{\mathbf{k}})^{2}}\\
 & +\hbar\frac{n_{0}T^{\mathrm{2B}}}{(\hbar\omega_{n})^{2}+(\hbar\omega_{\mathbf{k}})^{2}}=0,
\end{align*}
where the second term on both right-hand sides is the phase-fluctuation
propagator with $mc^{2}=n_{0}T^{\mathrm{2B}}$ and the dispersion
$\sqrt{2mc^{2}\epsilon_{\mathbf{k}}}$ is extended to the full Bogoliubov
dispersion $\hbar\omega_{\mathbf{k}}=\sqrt{\epsilon_{\mathbf{k}}\left(\epsilon_{\mathbf{k}}+2mc^{2}\right)}$.
In contrast to the exact phase-fluctuation propagator, the factor
$n_{c}/n$ is not present in the Bogoliubov propagator, which can
be attributed to a renormalization not present in Bogoliubov theory.
After the subtraction of the phase fluctuations, the propagator of
the non-phase fluctuations is given by
\begin{equation}
\hbar^{-1}\left\langle \phi'(\mathbf{k},\omega_{n})\phi'^{*}(\mathbf{k},\omega_{n})\right\rangle =\frac{i\hbar\omega_{n}+\epsilon_{\mathbf{k}}}{(\hbar\omega_{n})^{2}+(\hbar\omega_{\mathbf{k}})^{2}},\label{eq:High_En_fluc-1}
\end{equation}
while the anomalous averages vanish, i.e.,
\[
\left\langle \phi'(\mathbf{k},\omega_{n})\phi'(\mathbf{k},\omega_{n})\right\rangle =\left\langle \phi'^{*}(\mathbf{k},\omega_{n})\phi'^{*}(\mathbf{k},\omega_{n})\right\rangle =0.
\]
The vanishing of the anomalous averages means that the Green's function
is diagonal in Nambu space and this greatly simplifies the renormalization
procedure of the interaction.

\subsubsection*{Renormalization due to the non-phase fluctuations}

The accuracy of the action $S\left[n_{0},\theta,\phi'^{*},\phi'\right]$
can be improved systematically by incorporating the $\phi'$ fluctuations
into a renormalization of the action. However, due to the fundamental Ward identities associated with the $U(1)$ invariance of the theory, it turns out to be
more convenient to carry out this renormalization immediately at the level of
$S\left[\phi^{*},\phi\right]$, cf. Eq.~(\ref{eq:Action_Bose_Gas-1}),
and then apply the expansion of the field, as in Eq.~(\ref{eq:Expansion_of_field-1}). To be useful for a strong-coupling situation this renormalization should be carried out by a non-perturbative method, such as for instance the large-$N$ expansion or the renormalization group. We here discuss only the latter choice.
The exact Wilsonian renormalization-group flow equation for the action
$S\left[\phi^{*},\phi\right]$ is
\[
\frac{\mathrm{d}S}{\mathrm{d}\Lambda}=\frac{\hbar}{2}\mathrm{Tr}\left[\delta_{\Lambda}\ln\left(-\mathbf{G}'^{-1}+\frac{1}{\hbar}\frac{\delta^{2}S_{\mathrm{int}}}{\delta\mathbf{\Phi}\delta\mathbf{\Phi}^{*}}\right)\right],
\]
which is derived in Appendix \ref{sec:AP_RG}. Here $S\left[\phi^{*},\phi;\Lambda\right]$
is the effective action obtained by integrating out all non-phase
fluctuations above the momentum $\hbar\Lambda$, $\mathbf{G}'$ is
the matrix propagator of the non-phase fluctuations, $S{}_{\mathrm{int}}$
is the non-gaussian part of the effective action, the trace is over
space, imaginary time and Nambu space $\mathbf{\Phi}(\mathbf{k},\omega_{n})=\left[\phi'(\mathbf{k},\omega_{n}),\phi'^{*}(-\mathbf{k},-\omega_{n})\right]^{T}$,
and $\delta_{\Lambda}=\delta(k-\Lambda)$. Although there are no small
parameters in the theory of unitary Bose gases, the renormalization
group can distinguish between the relevance of the various coupling
constants based on their scaling dimension under renormalization.
As the effective interaction evaluated at zero momentum and zero frequency
is expected to be a crucial variable, since it induces a flow of the
chemical potential that corresponds to the most relevant operator
of the action, let us here restrict our attention to these parameters,
allowing us also to give an explicit illustration of the general procedure.
The running of the chemical potential and effective interaction $g$
are in general found to be given in terms of the so-called beta functions
by
\[
\Lambda\frac{\mathrm{d}\mu}{\mathrm{d}\Lambda}=\beta_{\mu}(\mu,g),\quad\Lambda\frac{\mathrm{d}g}{\mathrm{d}\Lambda}=\beta_{g}(\mu,g).
\]
By solving these equations the renormalized action $S\left[\phi^{*},\phi;\Lambda\right]$
is found. Then, after inserting the expansion of the field, the renormalized
action $S\left[n_{0},\theta,\phi'^{*},\phi';\Lambda\right]$ is obtained.
This action defines the propagator of the non-phase fluctuations in
terms of the effective interaction, which in this case is simply Eq.~(\ref{eq:High_En_fluc-1})
with the interaction replaced by the effective interaction at zero
momentum and zero frequency, namely $mc^{2}\equiv n_{0}g$. This can thus be seen as a self-consistency condition on the propagator of the non-phase fluctuations, which should be generalized when more running coupling constants are included.

Before we turn to the solution of the above renormalization-group equations,
we first show that our approach reproduces the exact propagator in
the long-wavelength limit derived by Nepomnyashchii and Nepomnyashchii,
as mentioned in the introduction, and that the condensate density
and the total density can in general be expressed in terms of the
quasi-condensate density and the effective interaction at zero frequency
and momentum.

\subsubsection*{Exact normal and anomalous propagators}

To reproduce the exact propagator in the long-wavelength limit we
take the Fourier transform of the exact one-particle correlation function,
which in our theory is given by
\begin{align*}
\left\langle \phi(\mathbf{x},\tau)\phi^{*}(\mathbf{0},0)\right\rangle  & =n_{0}\left\langle \exp\left[i\left(\theta(\mathbf{x},\tau)-\theta(\mathbf{0},0)\right)\right]\right\rangle \\
 & +\left\langle \phi'(\mathbf{x},\tau)\phi'^{*}(\mathbf{0},0)\right\rangle
\end{align*}
By expanding the exponential we find that the dominant long-wavelength
behavior is due only to the first three terms in the expansion, where
the first term is the condensate density and the second term is simply
the exact phase-fluctuation propagator in Eq.~(\ref{eq:Phase_fluc_prop-1}).
The third term gives a non-trivial logarithmic term, which results
from a convolution of two phase-fluctuation propagators as was shown
in section~\ref{sub:Difficulties_Bog}. The expansion is thus
\begin{align}
 & \hbar^{-1}\left\langle \phi(\mathbf{k},\omega_{n})\phi{}^{*}(\mathbf{k},\omega_{n})\right\rangle \nonumber \\
 & \simeq n_{c}\beta V\delta_{\mathbf{k},\mathbf{0}}\delta_{n,0}+\frac{\frac{n_{c}}{n}mc^{2}}{(\hbar\omega_{n})^{2}+2mc^{2}\epsilon_{\mathbf{k}}}\label{eq:Nepom_prop}\\
 & -\frac{3\sqrt{mc^{2}}}{32\sqrt{2}\epsilon_{F}^{3/2}}\frac{n_{c}}{n}\log\left[\frac{(\hbar\omega_{n})^{2}+2mc^{2}\epsilon_{\mathbf{k}}}{(8mc^{2})^{2}}\right].\nonumber
\end{align}
To obtain the denominator inside the logarithm using the phase-fluctuation
propagator in Eq.~(\ref{eq:Phase_fluc_prop-1}) with $mc^{2}=n_{0}g$,
an ultra-violet subtraction was needed. This subtraction removes the
ultra-violet divergences associated with a point interaction \citep{Stoof2009Ultracold},
and is a result of the renormalization of the bare coupling to $T^{\mathrm{2B}}$,
as explained in appendix \ref{sec:UVsubtr}. Also, it was used that
\begin{align*}
 & n_{0}\left\langle \exp\left[i\left(\theta(\mathbf{x},\tau)-\theta(\mathbf{0},0)\right)\right]\right\rangle \\
 & =n_{0}\exp\left[-\frac{1}{2}\left\langle \left[\theta(\mathbf{x},\tau)-\theta(\mathbf{0},0)\right]^{2}\right\rangle \right]\\
 & =n_{c}\exp\left[\left\langle \theta(\mathbf{x},\tau)\theta(\mathbf{0},0)\right\rangle \right],
\end{align*}
and the condensate density is defined in terms of the off-diagonal
long-range order of the one-particle density matrix
\begin{eqnarray}
n_{c} & \equiv & \lim_{|\mathbf{x}|\rightarrow\infty}\left\langle \phi(\mathbf{x},0)\phi^{*}(\mathbf{0},0)\right\rangle \label{eq:cond_dens}\\
 & = & n_{0}\exp\left[-\left\langle \theta(\mathbf{0},0)\theta(\mathbf{0},0)\right\rangle \right].\nonumber
\end{eqnarray}
In the last line it was used that in the limit of large separation $\left\langle \theta(\mathbf{x},0)\theta(\mathbf{0},0)\right\rangle =0$,
as is also shown in Appendix \ref{sec:UVsubtr}.$ $

Similarly, the exact anomalous propagator is given by
\begin{eqnarray*}
\left\langle \phi(\mathbf{x},\tau)\phi(\mathbf{0},0)\right\rangle  & = & n_{0}\left\langle \exp\left[i\left(\theta(\mathbf{x},\tau)-\theta(\mathbf{0},0)\right)\right]\right\rangle \\
 & = & n_{c}\exp\left[-\left\langle \theta(\mathbf{x},\tau)\theta(\mathbf{0},0)\right\rangle \right],
\end{eqnarray*}
such that the Fourier transform in the long-wavelength limit only
differs from Eq.~(\ref{eq:Nepom_prop}) by a minus sign in front
of the second term in the right-hand side. The above expressions are
the exact normal and anomalous propagators in the long-wavelength
limit, as derived in a different manner in Refs. \citep{Nepomnyashchii1975Contribution,Nepomnyashchii1978Infrared}.
In particular, this leads to the counter-intuitive conclusion that
the anomalous self-energy vanishes at zero momentum and zero frequency
\citep{Nepomnyashchii1978Infrared}.

\subsubsection*{Condensate density and total density}

The condensate density can be expressed in terms of the quasicondensate
density and the effective interaction using Eq.~(\ref{eq:cond_dens})
as
\begin{equation}
n_{c}=n_{0}\exp\left[\frac{3}{4}\left(2\sqrt{2}-\pi\right)\left(\frac{n_{0}g}{\epsilon_{F}}\right)^{3/2}\right],\label{eq:cond_density-1}
\end{equation}
In order to determine the condensate density, the quasicondensate
density $n_{0}$ needs to be eliminated in favor of the total density
$n=\left\langle \phi(\mathbf{x},\tau)\phi^{*}(\mathbf{x},\tau)\right\rangle $
using
\begin{equation}
n=n_{0}+\frac{1}{4}\left(8\sqrt{2}-3\pi\right)\left(\frac{n_{0}g}{\epsilon_{F}}\right)^{3/2}n,\label{eq:Total_density-1}
\end{equation}
where the second term is the contribution from the high-energy fluctuations
$n'=\left\langle \phi'(\mathbf{x},\tau)\phi'^{*}(\mathbf{x},\tau)\right\rangle $,
see Eq.~(\ref{eq:High_En_fluc-1}). As required, exactly the same
ultra-violet subtraction was used for the high-energy fluctuations
as in Eq.~(\ref{eq:cond_density-1}), see appendix \ref{sec:UVsubtr}.
To solve these equations only the effective interaction at zero momentum
and zero frequency remains to be determined using the renormalization
group.

To summarize thus our general approach,
the action $S\left[\phi{}^{*},\phi\right]$ of the Bose gas can be
systematically renormalized by the non-phase fluctuations $\phi'$
using for instance the renormalization-group flow equation, giving in particular
rise to an effective coupling $g$ and a renormalized chemical potential
$\mu$. The propagators of the non-phase fluctuations are determined
self-consistently after expansion of the field $\phi$. After this renormalization step has been performed, during which no infrared divergencies will occur, the exact propagator of the phase fluctuations can be used to reproduces the exact normal and anomalous propagators in the long-wavelength limit.

\subsection{Applications\label{sub:Results}}

In this section we apply this general framework within the simplest approximation that goes beyond the Bogoliubov theory to obtain several quantities of the Bose gas as a function of scattering length without encountering any infrared divergencies. We use this particular approximation mostly for illustrational purposes of the general procedure and as a proof of principle that in this manner finite results can be obtained even at unitarity. To formulate the most accurate approximation at unitarity is beyond the scope of this paper and is left for future work.

\subsubsection*{Bogoliubov theory revisited}

Within Bogoliubov theory the effective interection is assumed not to be running and we simply have that $g=g(\Lambda=0)=g(\Lambda=\infty)=T^{\mathrm{2B}}(-2n_{0}g)$ \citep{Stoof2009Ultracold}, where the energy dependence of the two-body T matrix is given by
\begin{equation}
T^{\mathrm{2B}}(E) = \frac{4\pi a \hbar^2}{m} \frac{1}{1-a \sqrt{-mE/\hbar^2}}.
\label{t2body}
\end{equation}
Note that in the boundary condition that is used, i.e., $g(\Lambda=\infty)=T^{\mathrm{2B}}(-2n_{0}g)$, the particular value of the energy argument of the two-body T matrix is such that indeed not only the dominant but also the subdominant ultra-violet term in $\beta_g$ is cancelled as shown explicitly in Eq.~(\ref{eq:Beta_func}) below. The chemical potential is running, however, with
\begin{equation}
\beta_{\mu} =-2g\frac{4\pi\Lambda^{3}}{(2\pi)^{3}}
  \left[\frac{\epsilon_{\Lambda}-\hbar\omega_{\Lambda}}{2\hbar\omega_{\Lambda}}
        +\frac{n_{0}g}{2\epsilon_{\Lambda}+2n_{0}g}\right],
\end{equation}
where the dispersions are evaluated at $\Lambda$. Integrating the resulting renormalization-group equation with the boundary condition $\mu(\Lambda=\infty)=gn_0$ gives ultimately
\begin{equation}
\mu = \mu(\Lambda=0) = (2n'+ n_0) T^{\mathrm{2B}}(-2n_{0}g),
\end{equation}
with $n' = n - n_0$ determined from Eq.~(\ref{eq:Total_density-1}). As desired, the latter equation exactly reproduces the chemical potential of the Bogoliubov theory, including the Lee-Huang-Yang correction. Furthermore, Eqs.~(\ref{eq:cond_density-1}) and (\ref{eq:Total_density-1}) also reproduce the condensate depletion of Eq.~(\ref{eq:depletion_condensate}) at weak coupling.

\subsubsection*{Effective interaction, (quasi-)condensate density and one-particle
density matrix}

To go beyond the Bogoliubov theory, we must now determine the effective interaction $g$ in a better approximation. Taking only the renormalization of the coupling
constant and the chemical potential into account, which we here use to illustrate the general procedure but interestingly enough turns out
to be very accurate for the unitary Fermi gas \citep{Gubbels2008Renormalization},
the beta functions are given by
\begin{figure}[b]
\centering{}\includegraphics[scale=0.35]{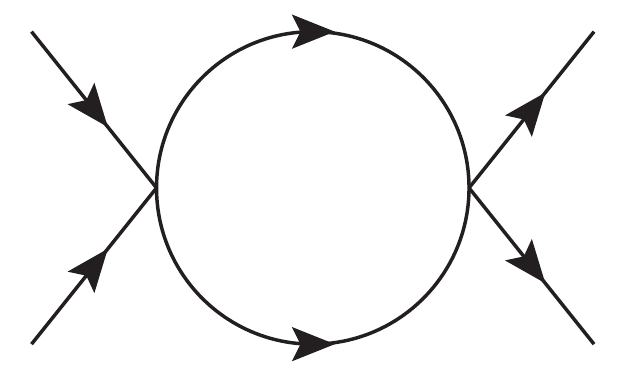}\qquad{}\qquad{}\includegraphics[scale=0.35]{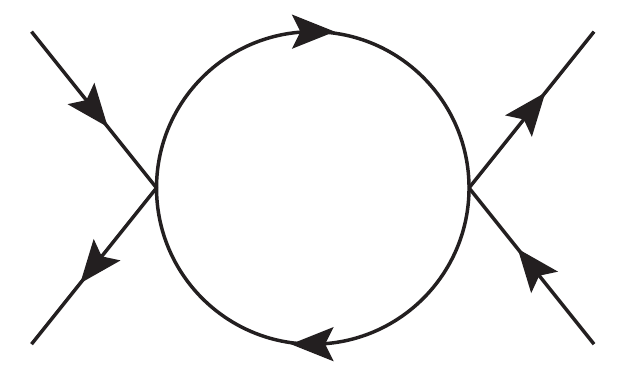}\caption{The Feynman diagrams for the fluctuations $\phi'$ that contribute
to the beta function $\beta_{g}$ and can be viewed as the ladder
sum $\Xi$ and bubble sum $\Pi$ contributions included in the Bethe-Salpeter
equation for the effective interaction $g$.\label{fig:loop_bubble_diagram}}
\end{figure}
\begin{align}
\beta_{\mu} & =-2g\frac{4\pi\Lambda^{3}}{(2\pi)^{3}}\left[v_{\Lambda}^{2}+\frac{n_{0}g}{2\epsilon_{\Lambda}+2n_{0}g}\right],\label{eq:Beta_func}\\
\beta_{g} & =g^{2}\frac{4\pi\Lambda^{3}}{(2\pi)^{3}}\left[\frac{u_{\Lambda}^{4}+v_{\Lambda}^{4}-8u_{\Lambda}^{2}v_{\Lambda}^{2}}{2\hbar\omega_{\Lambda}}-\frac{1}{2\epsilon_{\Lambda}+2n_{0}g}\right],\nonumber
\end{align}
where the Bogoliubov dispersion $\hbar\omega_{\mathbf{k}}$ and the
coherence factors $u_{\mathbf{k}}^{2}=v_{\mathbf{k}}^{2}+1=\left(\hbar\omega_{\mathbf{k}}+\epsilon_{\mathbf{k}}\right)/2\hbar\omega_{\mathbf{k}}$
are evaluated at $\Lambda$. For a derivation of these expressions,
compare with the frequency and momentum-dependent integral expressions
of Eqs.~(\ref{eq:loop}) and (\ref{eq:non_phas_fluc_UVsubtr}).

The effective interaction is obtained by integrating its differential
equation using the boundary condition
\[
\frac{1}{g(\Lambda=\infty)}=\frac{1}{T^{\mathrm{2B}}(-2n_{0}g)}=\frac{1}{T^{\mathrm{2B}}}-\frac{3\pi}{4\sqrt{2}}\sqrt{\frac{2n_{0}g}{\epsilon_{F}}}\frac{\epsilon_{F}}{n},
\]
where it must be noted that the effective interaction inside the Bogoliubov
dispersion is the fully renormalized value $g(\Lambda=0)$ which,
as previously explained, is determined self-consistently. Ultimately,
we obtain
\begin{eqnarray}
\frac{1}{g} & = & \frac{1}{T^{\mathrm{2B}}}-\left[\Xi\left(\mathbf{0},0\right)+4\Pi(\mathbf{0},0)\right]\label{eq:Eff_int2} \nonumber\\
 & = & \frac{1}{T^{\mathrm{2B}}}+\frac{1}{4\sqrt{2}\pi^{2}}\left(\frac{2m}{\hbar^{2}}\right)^{3/2}\sqrt{n_{0}g},
\end{eqnarray}
Note that this equation can also be obtained directly as the result
of a resummation of an infinite number of the diagrams shown in Fig.~\ref{fig:loop_bubble_diagram},
see appendix \ref{sec:AP_loop_bubble} for the derivation of the ladder
sum $\Xi$ and bubble sum $\Pi$ contributions. The equation also
shows the non-perturbative nature of the renormalization group.
\begin{figure}
\begin{centering}
\includegraphics[bb=10bp 10bp 350bp 220bp,clip,scale=0.7]{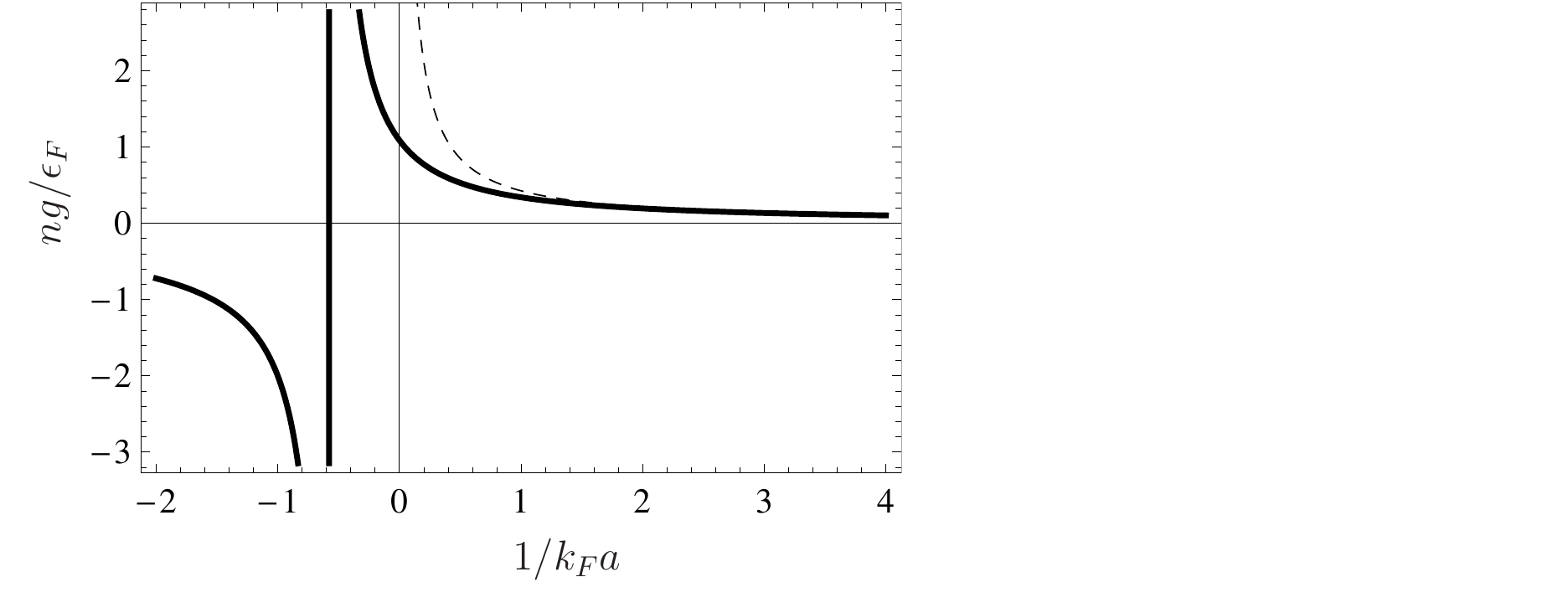}\caption{The effective interaction as a function of scattering length. The
gray dashed line is the weak-coupling limit for positive scattering
length.\label{fig:Eff_int}}
\includegraphics[bb=0bp 0bp 350bp 220bp,clip,scale=0.7]{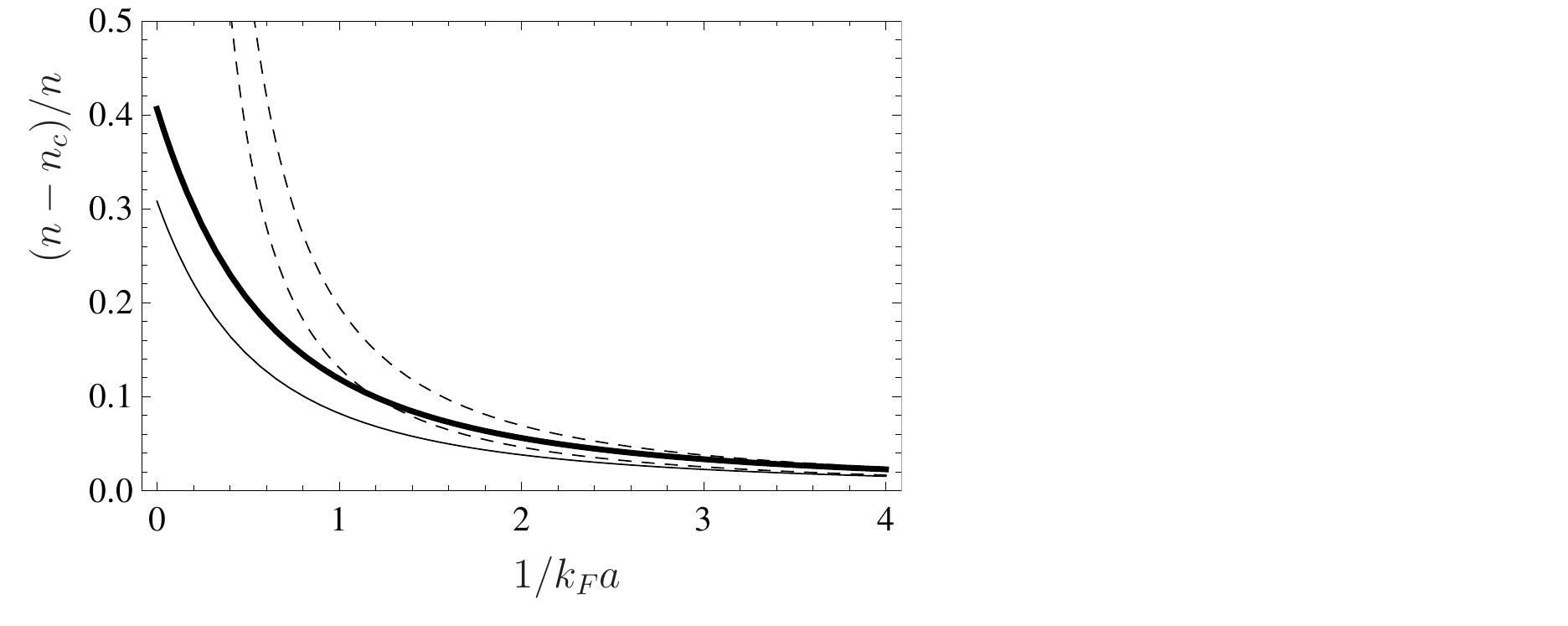}\caption{The fractional depletion from the condensate $(n-n_{c})/n$ (thick)
and from the quasi-condensate (thin) $n'/n=(n-n_{0})/n$ as a function
of scattering length. The dashed lines are the weak-coupling result
for the condensate density in Eq.~(\ref{eq:depletion_condensate})
and for the quasi-condensate density as derived in Ref. \citep{Stoof2009Ultracold}.\label{fig:Cond_dens}}

\par\end{centering}

\begin{centering}
\includegraphics[bb=10bp 10bp 365bp 226bp,clip,scale=0.68]{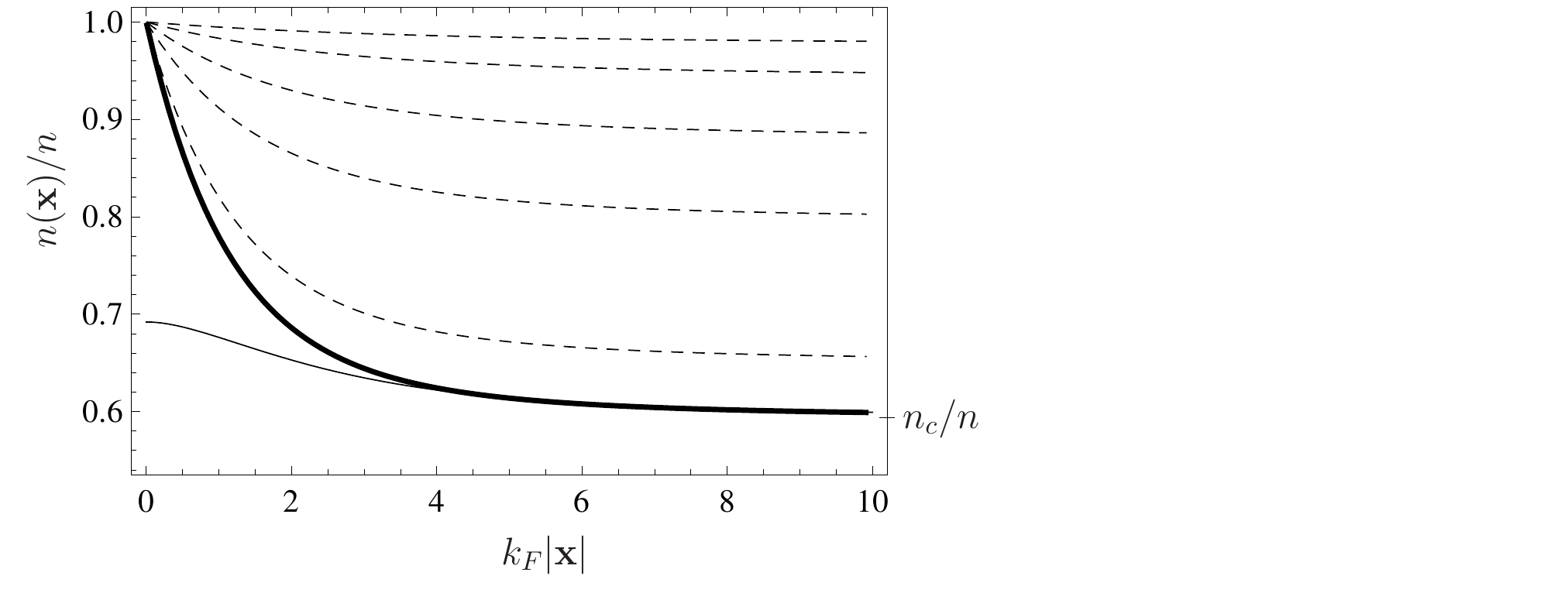}
\par\end{centering}

\caption{The one-particle density matrix $n(\mathbf{x})/n$ as a function $k_{F}|\mathbf{x}|$.
Here the thick line is the one-particle density matrix at unitarity
with the thin line its contributions due to the condensate and its
phase fluctuations. From the difference of the two graphs the contribution
coming from non-phase fluctuations can be inferred. The condensate
density $n_{c}/n$ at unitarity is indicated on the right. The dashed
lines correspond to the one-particle density matrix with a finite
scattering length, namely with $1/k_{F}a=\left\{ 1/10,1/2,1,2,4\right\} $
from bottom to top, respectively.\label{fig:One_part_dens_matrix}}
\end{figure}

The effective interaction and the condensate density in terms of the
total density are found analytically as a function of scattering length
by solving Eqs.~(\ref{eq:cond_density-1}), (\ref{eq:Total_density-1}), and (\ref{eq:Eff_int2}), and are
plotted in Figs. \ref{fig:Eff_int} and \ref{fig:Cond_dens}. As can
be seen from Fig.~\ref{fig:Eff_int} the position of the resonance
shifts due to many-body effects to negative scattering lengths as
a consequence of the screening effects of the bubble sum. In the unitarity
limit, $T^{2B}\rightarrow\infty$, the effective interaction and condensate
density are in this first approximation given by
\begin{align*}
\frac{ng}{\epsilon_{F}} & =\frac{2}{3^{2/3}}\left(1+\lambda\right)^{1/3}\simeq1.09,\\
\frac{n_{c}}{n} & =\frac{1}{1+\lambda}\exp\left(\frac{2\sqrt{2}-\pi}{\sqrt{2}(1+\lambda)}\right)\simeq0.59,\\
\frac{n'}{n} & =\frac{n-n_{0}}{n}=\frac{\lambda}{1+\lambda}\simeq0.31,
\end{align*}
where
\[
\lambda\equiv\frac{n'}{n_{0}}=\frac{1}{3\sqrt{2}}\left(8\sqrt{2}-3\pi\right)\simeq0.45.
\]
The depletion from the condensate is given by $1-n_{c}/n\simeq0.41$,
which clearly differs from the density of particles contributing to
the non-phase fluctuating modes $n'$ by phase-fluctuation contributions.

The one-particle density matrix is defined by
\begin{eqnarray*}
n(\mathbf{x}) & = & \left\langle \phi(\mathbf{x},0)\phi^{*}(\mathbf{0},0)\right\rangle \\
 & = & n_{c}\exp\left[\left\langle \theta(\mathbf{x},0)\theta(0,0)\right\rangle \right]+\left\langle \phi'(\mathbf{x},0)\phi'^{*}(\mathbf{0},0)\right\rangle ,
\end{eqnarray*}
where the expressions of the phase-fluctuation and the non-phase fluctuation
propagator with the appropriate ultra-violet subtractions can be found
in appendix \ref{sec:UVsubtr}. The one-particle density matrix is
shown for several scattering lengths in Fig.~\ref{fig:One_part_dens_matrix},
including at unitarity. Clearly the condensate density reduces to
the total density in the weak-coupling limit.

\subsubsection*{Chemical potential and speed of sound}

The change in the chemical potential follows from integrating Eq.~(\ref{eq:Beta_func})
and is given by $\Delta\mu=2n'g$. According to the exact Hugenholtz-Pines
theorem \citep{Hugenholtz1959GroundState} the chemical potential
in our theory is then given by $\mu=n_{0}g+\Delta\mu=n_{0}g+2n'g$.
The value of the chemical potential at unitarity is found to be
\[
\frac{\mu}{\epsilon_{F}}=\frac{n_{0}g+2n'g}{\epsilon_{F}}=\frac{2}{3^{2/3}}\frac{1+2\lambda}{\left(1+\lambda\right)^{2/3}}\simeq1.42.
\]
The chemical potential at unitarity is usually written as $\mu=(1+\beta)\epsilon_{F}$,
such that we have for the universal constant $\beta\simeq0.42$. Furthermore,
the speed of sound at unitarity is given by
\[
\frac{mc^{2}}{\epsilon_{F}}=\frac{n_{0}g}{\epsilon_{F}}=\frac{1}{1+2\lambda}\frac{\mu}{\epsilon_{F}}\simeq0.53\frac{\mu}{\epsilon_{F}}\simeq0.75.
\]
The expected value for the speed of sound at unitarity in terms of
the chemical potential is $mc^{2}=n(\mathrm{d}\mu/\mathrm{d}n)=2\mu/3\simeq0.66\mu$,
which is close to our result and gives an indication of the accuracy of the simplest
first approximation that we have presented here.

In comparison to the literature, our results for the chemical potential
differ from the variational studies which find $\beta\simeq-0.2$
\citep{Song2009Ground} and $\beta\simeq1.93$ \citep{Cowell2002Cold}
and the renormalization-group study ($\beta\simeq-0.34$) \citep{Lee2010Universality}.
As mentioned in the introduction, it is not clear that the variational
studies are always inside a Hilbert space orthogonal to the true many-body
ground state. Also, as correctly presented in these articles, these
variational results should not be viewed as upper bounds to $\beta$,
as it is the energy which is determined variationally and not its
derivative with respect to the number of atoms. Furthermore, the variational
study in Ref.~\citep{Song2009Ground} always has an attractive interaction
whose normal mean-field contribution is treated in the Hartree-Fock
approximation, which presumably explains its negative value of $\beta$.
In contrast, our result uses for both the normal and anomalous contributions
an effectively repulsive interaction and as a result $\beta$ becomes
positive.

\subsubsection*{Contact}

Another interesting property is called the contact $C$ and is related
to the short-wavelength behavior of the momentum distribution, namely
\citep{Tan2008Large,Braaten2011Universal}
\[
n(\mathbf{k})\simeq C/\mathbf{k}^{4}.
\]
The value of the contact is determined by the non-phase fluctuations
and is found after performing the Matsubara sum over Eq.~(\ref{eq:High_En_fluc-1})
and expanding for large momenta to be
\begin{equation}
\frac{C}{k_{F}^{4}}=\left(\frac{n_{0}g}{2\epsilon_{F}}\right)^{2}.\label{eq:contact}
\end{equation}
This expression is of the same form as that found in Bogoliubov theory
\citep{Schakel2010Tan} but with the two-body $T$ matrix replaced
by the effective interaction. At unitarity, its value is
\[
\frac{C}{k_{F}^{4}}=\frac{1}{3^{4/3}}\frac{1}{\left(1+\lambda\right)^{4/3}}\simeq0.14
\]

An equivalent definition of the contact is through the average of
the interaction term in the action \citep{Schakel2010Tan,Braaten2008Exact,Braaten2011Universal}
\begin{equation}
\frac{C}{k_{F}^{4}}=\left(\frac{T^{\mathrm{2B}}}{2\epsilon_{F}}\right)^{2}\left\langle \left|\phi\right|^{4}\right\rangle .\label{eq:contact_average}
\end{equation}
Assuming that the action is first renormalized, such that the two-body
$T$-matrix is replaced by the effective interaction $g$, and that
all non-phase fluctuations have been included into the renormalization
of the action, i.e., we take $\left\langle |\phi|^{4}\right\rangle =n_{0}^{2}$
to avoid double counting, we re-obtain Eq.~(\ref{eq:contact}).

Yet another definition of the contact can be given in terms of the
derivative of the total energy or the chemical potential with respect
to the scattering length, namely
\begin{eqnarray}
\frac{C}{k_{F}^{4}} & = & -\frac{4\pi}{\epsilon_{F}k_{F}^{4}}\frac{\mathrm{d}(E/V)}{\mathrm{d}(1/a)},\label{eq:contact_mu}
\end{eqnarray}
where the total energy per volume is obtained from the chemical potential
as $E/V=\int_{0}^{n}\mu(n',a)\mathrm{d}n'$. By neglecting the contribution
of the non-phase fluctuations in the chemical potential, i.e., taking
$\mu=n_{0}g$, we analytically re-obtain the same value of the contact
at unitarity as obtained from Eq.~(\ref{eq:contact}) and numerically
we re-obtain the same contact as a function of scattering length.
If the contact is determined through the derivative of the complete
chemical potential, which includes
contributions from the non-phase fluctuations, it becomes larger.
We expect that this difference is a consequence of a double counting,
since the effects of non-phase fluctuations have already been included
in the effective interaction and should not be included again through
the derivative of the self-energy contribution $2n'g$ of the chemical
potential.

\subsubsection*{Energy-dependent effective interaction and bound state}

\begin{figure}[t]
\begin{centering}
\includegraphics[bb=10bp 5bp 320bp 208bp,clip,scale=0.7]{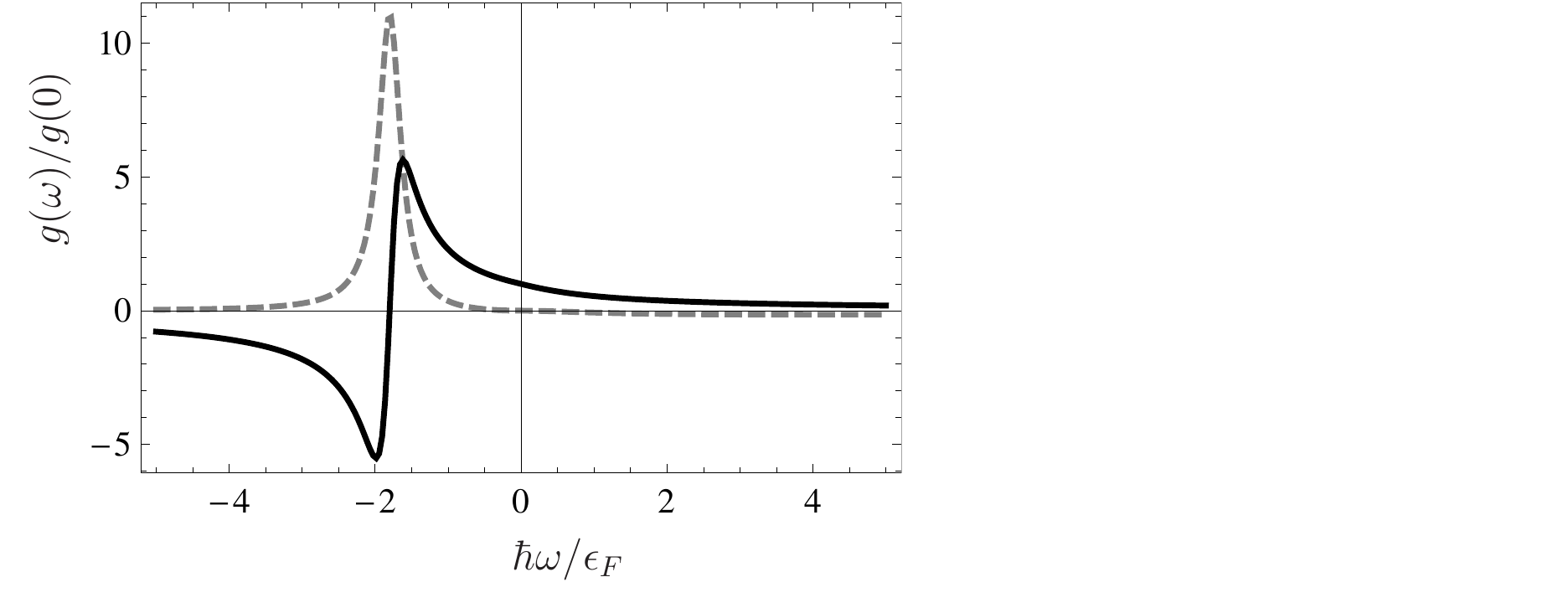}
\par\end{centering}

\begin{centering}
\caption{The real (solid) and imaginary (dashed) part of the effective interaction
at unitarity normalized to its value at zero frequency as a function
of center-of-mass frequency.\label{fig:Freqdep_eff_int}}

\includegraphics[bb=5bp 10bp 310bp 210bp,clip,scale=0.7]{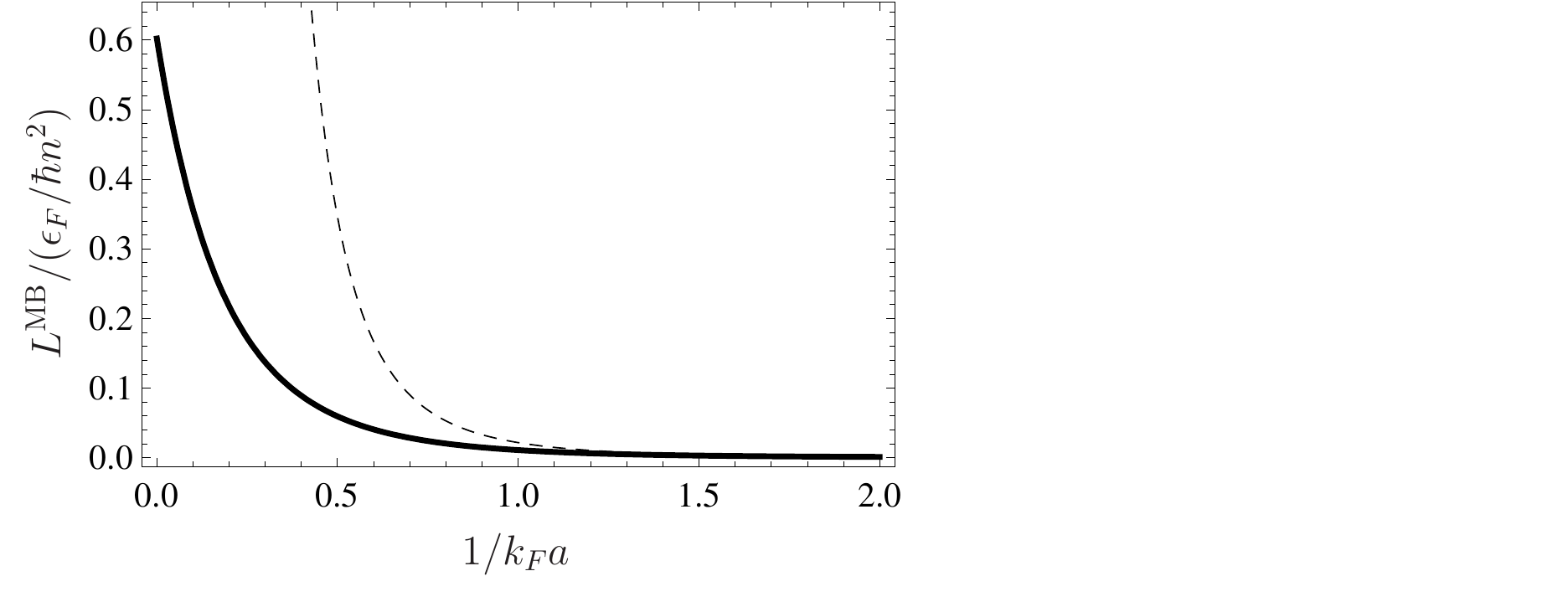}
\par
\end{centering}

\caption{The many-body recombination rate as a function of scattering length.
The thick line is determined using the contact in Eq.~(\ref{eq:contact}).
The three-body recombination rate in Eq.~(\ref{eq:L3B}) is shown
as the dashed line.\label{fig:LMB}}
\end{figure}

The center-of-mass energy dependence is most easily investigated by
generalizing Eq.~(\ref{eq:Eff_int2}) for non-zero frequencies, giving
\[
\frac{1}{g(\hbar\omega_{n})}=\frac{1}{T^{\mathrm{2B}}}-\left[\Xi\left(\mathbf{0},\omega_{n}\right)+4\Pi(\mathbf{0},0)\right].
\]
Here the bubble sum contribution $\Pi$ is not energy dependent, since
at this level of approximation it only depends on the relative energy.
The frequency-dependent ladder contribution can be found analytically
and its integral expression is shown in appendix \ref{sec:AP_loop_bubble}.
For high energies the expression reduces to the vacuum expression
in Eq.~(\ref{t2body}). The frequency dependence of
the effective interaction at unitarity is shown in Fig.~\ref{fig:Freqdep_eff_int},
where the Kramers-Kronig-like feature in the real and imaginary parts,
that is a consequence of the molecular bound state, is clearly visible.
This feature shifts to more negative frequencies for decreasing scattering
lengths.

\subsubsection*{Three-body recombination rate}

As mentioned in the introduction, the atomic Bose gas is meta-stable.
The primary mechanism for the system to decay to the true ground state
of a Bose-Einstein condensate of molecules is by inelastic three-body
collisions. In these collisions three particles interact to form a
diatomic molecule and a free atom. The molecular binding energy is
then released in the form of kinetic energy of the molecule and atom,
which results in a loss of atoms from the shallow traps used in cold
atomic gas experiments. Here the dependence of the decay rate on the
scattering length is investigated using our knowledge of the contact
and of how the bound-state energy is shifted away from the original
position of the resonance due to many-body effects.

The particle loss is written as
\[
\frac{\mathrm{d}n}{\mathrm{d}t}=-Ln^{3},
\]
where $L$ is the three-body loss rate \citep{Braaten2006Universality,Diederix2011Ground}.
The dependence on the scattering length of the loss rate is found
by application of Fermi's golden rule
\[
L\propto\left|\left\langle f\right|V\left|i\right\rangle \right|^{2}q_{f}.
\]
Here $\left|f\right\rangle $ and $\left|i\right\rangle $ indicate
the final and initial state, respectively, and $q_{f}$ is the wavevector
of the final state. The final state is the Feshbach bound state \citep{Stoof2009Ultracold}
and is given by
\[
\left\langle \mathbf{r}|f\right\rangle =\frac{1}{\sqrt{2\pi a_{\mathrm{b}}}}\frac{e^{-r/a_{\mathrm{b}}}}{r},
\]
where we defined the effective scattering length $a_{\mathrm{b}}$
that without many-body corrections is just equal to $a(B)$. The wavevector
of the final state is given by $q_{f}\propto a_{\mathrm{b}}^{-1}$.
The two-body scattering states in the open channel \citep{Stoof2009Ultracold}
are given by
\[
\lim_{r\downarrow0}\left\langle \mathbf{r}|\psi^{(+)}(\mathbf{k})\right\rangle \simeq1-\frac{a}{r}.
\]
The initial state can be viewed as a product of three such scattering
states, such that for small radii where the interaction potential
is non-vanishing
\[
\left\langle \mathbf{r}_{12}\mathbf{r}_{23}|i\right\rangle \propto a^{3}.
\]
When no many-body corrections are present, we therefore expect
\[
L^{\mathrm{3B}}(a)\propto\left(\frac{1}{\sqrt{a_{\mathrm{b}}}}a^{3}\right)^{2}\frac{1}{a_{\mathrm{b}}}\propto\frac{a^{6}}{a_{\mathrm{b}}^{2}}\propto a^{4}.
\]
From Efimov physics it is known that for a shallow bound state
\begin{equation}
L^{\mathrm{3B}}(a)=F(a)\frac{\hbar}{2m}a^{4},\label{eq:L3B}
\end{equation}
where $F(a)$ is a logarithmically periodic function of the scattering
length and its maximum value is $F_{\mathrm{max}}\simeq67.12$ \citep{Braaten2006Universality,Zaccanti2009Observation,Pollack2009Universality}.
From now on we neglect the Efimov physics and concentrate on the maximum
value $L^{\mathrm{3B}}(a)=F_{\mathrm{max}}\hbar a^{4}/2m$.

When the scattering length becomes large, many-body effects become
important. The scattering state is then renormalized by the wavefunction
renormalization factor $\sqrt{Z(a)}$, which leads to the renormalized
initial state
\[
\left\langle \mathbf{r}_{12}\mathbf{r}_{23}|i\right\rangle \propto(\sqrt{Z}a)^{3}=\frac{C^{3/2}}{(4\pi n)^{3}},
\]
where it was used that the wavefunction renormalization factor can
be related to the contact by $C=Z(4\pi an)^{2}$ \citep{Diederix2011Ground}.
The effective scattering length $a_{\mathrm{b}}$ is given in terms
of the bound-state energy $E_{b}(a)=-\hbar^{2}/ma_{\mathrm{b}}^{2}(a)$.
The many-body loss rate can then be expressed in terms of the contact
and the bound-state energy, namely
\begin{eqnarray*}
L^{\mathrm{MB}}(a) & = & \left(F_{\mathrm{max}}\frac{\hbar}{2m}\right)\left[\frac{1}{\sqrt{a_{\mathrm{b}}}}\frac{C^{3/2}}{(4\pi n)^{3}}\right]^{2}\frac{1}{a_{\mathrm{b}}}\\
 & = & -F_{\mathrm{max}}\frac{1}{2}\frac{C^{3}(a)E_{\mathrm{b}}(a)}{\hbar(4\pi n)^{6}}.
\end{eqnarray*}
In dimensionless form the many-body recombination rate is
\begin{eqnarray*}
L^{\mathrm{MB}}/\left(\frac{\epsilon_{F}}{\hbar n^{2}}\right) & = & -F_{\mathrm{max}}\frac{\pi^{2}}{2}\left(\frac{3}{4}\right)^{4}\left(\frac{C}{k_{F}^{4}}\right)^{3}\frac{E_{\mathrm{b}}}{\epsilon_{F}},
\end{eqnarray*}
Here the last line is found using Eq.~(\ref{eq:contact}) for the
contact. The many-body recombination rate as a function of scattering
length is shown in Fig.~\ref{fig:LMB}.

At unitarity, where the bound-state energy is $E_{\mathrm{b}}\simeq-2.39n_{0}g\simeq-1.80\epsilon_{F}$, see Fig.~\ref{fig:Freqdep_eff_int},
this gives for the universal recombination rate
\[
L^{\mathrm{MB}}/\left(\frac{\epsilon_{F}}{\hbar n^{2}}\right)\simeq\frac{\pi^{2}\left(2.39F_{\mathrm{max}}\right)}{2^{8}3^{2/3}(1+\lambda)^{14/3}}\simeq0.61.
\]
The dependence of $L^{\mathrm{MB}}$ indicates that the many-body
loss rate saturates at unitarity to a finite value. A similar saturation
of the loss rate was seen experimentally in non-degenerate Bose gases
at unitarity in Ref. \citep{Rem2013Lifetime}, where the saturation
is determined by the temperature. When the temperature becomes small
the many-body loss rate is eventually set only by the density and
this crossover is determined by a universal function of $k_{B}T/\epsilon_{F}$ \cite{Zoran2013}.

\section{Discussion and conclusions\label{concl}}

Due to the fact that we have for illustrative purposes made the simplyfing assumption of having only two running coupling constants, all quantities in this article have been determined analytically as a function of scattering length, which allows us to compare to the known weak-coupling results, of which some are shown in section~\ref{sub:Bose_gas}.
Furthermore, we have taken the fully-renormalized value of the effective
interaction inside the Bogoliubov dispersions of the renormalization-group
flow equations, in Eq.~(\ref{eq:Beta_func}). Therefore, it would
be interesting to see the effect of a full numerical solution of the
coupled renormalization-group-flow equations, which is also of interest
for a study of the stability of the present results. The latter is also true for the study of the effects of various other coupling constants. Two important effects immediately come to mind. Due to the presence of a Feshbach bound state, the energy dependence of the effective interaction may play an important role. In addition, an important feature of the Bose gases near a Feshbach resonance is Efimov physics, which can also be studied by renormalization-group methods \cite{rg1,rg2}. By including the running of the appropriate three-body coupling constants in the renormalization-group equation it may be possible to investigate how much of the Efimov physics survives in a many-body setting when also medium effects are playing an important role. Another useful direction is to
obtain the renormalized thermodynamic potential of the theory. This
will allow for the determination of all quantities using thermodynamic
relations.

In summary, we have constructed a general self-consistent approach to describe strongly interacting Bose gases as a function of scattering length, which is free of infrared divergencies, can be improved systematically by renormalization-group methods or other non-pertubative methods, and reduces to the Bogoliubov theory for small scattering lengths. The generalization of the theory to non-zero temperature is straightforward,
see appendix \ref{sec:AP_loop_bubble}. Furthermore, we expect that
the approach can be applied to other systems with a broken continuous
symmetry, where similar infrared divergencies occur as a consequence
of the presence of Goldstone modes. We hope that our results stimulate
further experimental developments toward unitarity-limited Bose gases
in the near future.

\begin{acknowledgments}
This work is supported by the Stichting voor Fundamenteel Onderzoek
der Materie (FOM) and the Nederlandse Organisatie voor Wetenschaplijk
Onderzoek (NWO).
\end{acknowledgments}

\appendix

\section{Ladder and bubble-sum contributions\label{sec:AP_loop_bubble}}

In this section the ladder and bubble-sum contributions to the effective
interaction are derived. The full energy-momentum and temperature-dependent
ladder contribution is
\begin{widetext}
\begin{align}
\Xi(\mathbf{k},\omega_{n}) & =\frac{1}{\hbar^{2}\beta V}\sum_{\mathbf{k}',n'}G'(\mathbf{k}'_{+},\omega_{n'_{+}})G'(\mathbf{k}'_{-},\omega_{n'_{-}})\label{eq:full_loop}\\
 & =\frac{1}{V}\sum_{\mathbf{k}'}\left\{ \frac{u_{\mathbf{k}'_{+}}^{2}u_{\mathbf{k}'_{-}}^{2}}{i\hbar\omega_{n}-\hbar\omega_{\mathbf{k}'_{+}}-\hbar\omega_{\mathbf{k}'_{-}}}\left(\left[1+N(\hbar\omega_{\mathbf{k}'_{+}})\right]\left[1+N(\hbar\omega_{\mathbf{k}'_{-}})\right]-N(\hbar\omega_{\mathbf{k}'_{+}})N(\hbar\omega_{\mathbf{k}'_{-}})\right)\right.\nonumber \\
 & -\frac{u_{\mathbf{k}'_{+}}^{2}v_{\mathbf{k}'_{-}}^{2}}{i\hbar\omega_{n}-\hbar\omega_{\mathbf{k}'_{+}}+\hbar\omega_{\mathbf{k}'_{-}}}\left(N(\hbar\omega_{\mathbf{k}'_{+}})\left[1+N(\hbar\omega_{\mathbf{k}'_{-}})\right]-\left[1+N(\hbar\omega_{\mathbf{k}'_{+}})\right]N(\hbar\omega_{\mathbf{k}'_{-}})\right)\nonumber \\
 & -\frac{v_{\mathbf{k}'_{+}}^{2}u_{\mathbf{k}'_{-}}^{2}}{i\hbar\omega_{n}+\hbar\omega_{\mathbf{k}'_{+}}-\hbar\omega_{\mathbf{k}'_{-}}}\left(\left[1+N(\hbar\omega_{\mathbf{k}'_{+}})\right]N(\hbar\omega_{\mathbf{k}'_{-}})-N(\hbar\omega_{\mathbf{k}'_{+}})\left[1+N(\hbar\omega_{\mathbf{k}'_{-}})\right]\right)\nonumber \\
 & \left.+\frac{v_{\mathbf{k}'_{+}}^{2}v_{\mathbf{k}'_{-}}^{2}}{i\hbar\omega_{n}+\hbar\omega_{\mathbf{k}'_{+}}+\hbar\omega_{\mathbf{k}'_{-}}}\left(N(\hbar\omega_{\mathbf{k}'_{+}})N(\hbar\omega_{\mathbf{k}'_{-}})-\left[1+N(\hbar\omega_{\mathbf{k}'_{+}})\right]\left[1+N(\hbar\omega_{\mathbf{k}'_{-}})\right]\right)\right\} ,\nonumber
\end{align}

\end{widetext}
where we defined $\mathbf{k}'_{\pm}=\mathbf{k}/2\pm\mathbf{k}'$,
$n'_{\pm}=n/2\pm n'$, $(\hbar\omega_{\mathbf{k}})^{2}=\epsilon_{\mathbf{k}}(\epsilon_{\mathbf{k}}+2mc^{2})$
and the coherence factors $u_{\mathbf{k}}^{2}=v_{\mathbf{k}}^{2}+1=\left(\hbar\omega_{\mathbf{k}}+\epsilon_{\mathbf{k}}\right)/2\hbar\omega_{\mathbf{k}}$.
The bubble diagram is given by

\begin{align*}
\Pi(\mathbf{k},\omega_{n}) & =\frac{1}{\hbar^{2}\beta V}\sum_{\mathbf{k}',n'}G'(\mathbf{k}'_{+},\omega_{n'_{+}})G'(-\mathbf{k}'_{-},-\omega_{n'_{-}}),
\end{align*}
whose form is obtained from Eq.~(\ref{eq:full_loop}) by substituting
$u_{\mathbf{k}'_{-}}^{2}\leftrightarrow-v_{\mathbf{k}'_{-}}^{2}$,
while not modifying $u_{\mathbf{k}'_{+}}^{2},v_{\mathbf{k}'_{+}}^{2}$
, in the expression for the ladder contribution. In the zero-temperature
limit only the first and last line of both the ladder and bubble sum
contribution survives, where the former is
\begin{eqnarray}
\Xi(\mathbf{k},\omega_{n}) & = & \frac{1}{V}\sum_{\mathbf{k}'}\left[\frac{u_{\mathbf{k}'_{+}}^{2}u_{\mathbf{k}'_{-}}^{2}}{i\hbar\omega_{n}-\hbar\omega_{\mathbf{k}'_{+}}-\hbar\omega_{\mathbf{k}'_{-}}}\right.\label{eq:loop}\\
 &  & \hphantom{\frac{1}{V}\sum_{\mathbf{k}'}}\left.-\frac{v_{\mathbf{k}'_{+}}^{2}v_{\mathbf{k}'_{-}}^{2}}{i\hbar\omega_{n}+\hbar\omega_{\mathbf{k}'_{+}}+\hbar\omega_{\mathbf{k}'_{-}}}\right].\nonumber
\end{eqnarray}
In cold atomic gases the momentum dependence of these quantities is
of little importance. The momentum-independent ladder and bubble sum
contributions can then be integrated analytically, however, due to
the size of the expressions they are not shown here. Evaluating the
expressions also at zero frequency we obtain
\begin{eqnarray*}
\Xi(\mathbf{0},0) & = & \frac{3}{4\sqrt{2}\pi^{2}}\sqrt{\frac{n_{0}g}{\epsilon_{F}}}\frac{k_{F}^{3}}{\epsilon_{F}},\\
\Pi(\mathbf{0},0) & = & -\frac{1}{4\sqrt{2}\pi^{2}}\sqrt{\frac{n_{0}g}{\epsilon_{F}}}\frac{k_{F}^{3}}{\epsilon_{F}}.
\end{eqnarray*}
For the non-interacting case, where $\hbar\omega_{\mathbf{k}}=\epsilon_{\mathbf{k}}$
and $u_{\mathbf{k}}^{2}=v_{\mathbf{k}}^{2}+1=1$, we have that the
bubble sum contribution vanishes and that at zero momentum the ladder
contribution becomes
\begin{align}
\Xi(\mathbf{0},z) & =\frac{1}{V}\sum_{\mathbf{k}'}\left(\frac{1}{z-2\epsilon_{\mathbf{k}}}+\frac{1}{2\epsilon_{\mathbf{k}}}\right)\nonumber \\
 & =\frac{1}{8\sqrt{2}\pi}\left(\frac{2m}{\hbar^{2}}\right)^{3/2}\sqrt{-z},\label{eq:vacuum_loop}
\end{align}
where $z=i\hbar\omega_{n}$ and an ultra-violet subtraction was
needed as a consequence of the point interaction and in agreement with Eq.~(\ref{t2body}).

\section{Renormalization group\label{sec:AP_RG}}

Here we derive the renormalization-group flow equation. Starting from
the action of a homogeneous Bose gas as shown in Eq.~(\ref{eq:Action_Bose_Gas-1}),
we take the Fourier transform of the fields
\begin{equation}
\phi(\mathbf{x},\tau)=\frac{1}{\sqrt{\hbar\beta V}}\sum_{n}\sum_{\mathbf{k}<\Lambda}\phi_{\mathbf{k},n}e^{i(\mathbf{k}\cdot\mathbf{x}-\omega_{n}\tau)}
\end{equation}
and split up the field in terms of low-momentum $\phi_{<}$ and high-momentum
modes $\phi_{>}$ as
\begin{equation}
\phi(\mathbf{x},\tau)=\phi_{<}(\mathbf{x},\tau)+\phi_{>}(\mathbf{x},\tau),
\end{equation}
where the low-momentum and high-momentum modes are defined as
\begin{align*}
\phi_{<}(\mathbf{x},\tau) & =\frac{1}{\sqrt{\hbar\beta V}}\sum_{n}\sum_{\mathbf{k}<\Lambda}\phi_{\mathbf{k},n}e^{i(\mathbf{k}\cdot\mathbf{x}-\omega_{n}\tau)},\\
\phi_{>}(\mathbf{x},\tau) & =\frac{1}{\sqrt{\hbar\beta V}}\sum_{n}\sum_{\Lambda<\mathbf{k}<\Lambda+\mathrm{d}\Lambda}\phi_{\mathbf{k},n}e^{i(\mathbf{k}\cdot\mathbf{x}-\omega_{n}\tau)}.
\end{align*}
Rewriting the partition function leads to
\begin{align*}
 & Z=\int\mathcal{D}\phi^{*}\mathcal{D}\phi^{*}\exp\left\{ -\hbar^{-1}\left(S_{0}\left[\phi^{*},\phi\right]+S_{\mathrm{int}}\left[\phi^{*},\phi\right]\right)\right\} \\
 & =\int\mathcal{D}\phi_{<}^{*}\mathcal{D}\phi_{<}^{*}\exp\left\{ -\hbar^{-1}\left(S_{0}\left[\phi_{<}^{*},\phi_{<}\right]+S_{\mathrm{int}}\left[\phi_{<}^{*},\phi_{<}\right]\right)\right\} \\
 & \times\left[\int\mathcal{D}\phi_{>}^{*}\mathcal{D}\phi_{>}^{*}\exp\left\{ -\hbar^{-1}S_{0}\left[\phi_{>}^{*},\phi_{>}\right]\right\} \right.\\
 & \hphantom{\times}\left.\times\exp\left\{ -\hbar^{-1}\left(S_{\mathrm{int}}\left[\phi^{*},\phi\right]-S_{\mathrm{int}}\left[\phi_{<}^{*},\phi_{<}\right]\right)\right\} \vphantom{\int}\right],
\end{align*}
where the gaussian part of the action is denoted by $S_{0}\left[\phi^{*},\phi\right]$
and the non-gaussian part by $S_{\mathrm{int}}\left[\phi^{*},\phi\right]$.
Expanding up to second order in the high-momentum fields gives
\begin{align*}
 & -\hbar^{-1}S_{0}\left[\phi_{>}^{*},\phi_{>}\right]-\hbar^{-1}\left(S_{\mathrm{int}}\left[\phi^{*},\phi\right]-S_{\mathrm{int}}\left[\phi_{<}^{*},\phi_{<}\right]\right)\\
 & =-\frac{1}{2}\mathrm{Tr}\left[\mathbf{\Phi}_{>}^{\dagger}\left[-G_{0}^{-1}+\frac{1}{\hbar}\frac{\delta S_{\mathrm{int}}}{\delta\mathbf{\Phi}\delta\mathbf{\Phi}^{\dagger}}\left[\phi_{<}^{*},\phi_{<}^{*}\right]\right]_{\phi_{>}=0}\mathbf{\Phi}_{>}\right].
\end{align*}
Here the trace is over momentum, frequency and Nambu space $\mathbf{\Phi}(\mathbf{k},\omega_{n})=\left[\phi(\mathbf{k},\omega_{n}),\phi^{*}(-\mathbf{k},-\omega_{n})\right]^{T}$.
By integrating out the high-momentum fields, we obtain the effective
action for the low-momentum fields
\begin{align*}
-\hbar^{-1}S\left[\phi_{<}^{*},\phi_{<}\right] & =-\hbar^{-1}S_{0}\left[\phi_{<}^{*},\phi_{<}\right]-\hbar^{-1}S_{\mathrm{int}}\left[\phi_{<}^{*},\phi_{<}\right]\\
 & -\frac{1}{2}\mathrm{Tr}\ln\left[-G_{0}^{-1}+\frac{1}{\hbar}\frac{\delta S_{\mathrm{int}}}{\delta\mathbf{\Phi}\delta\mathbf{\Phi}^{\dagger}}\left[\phi_{<}^{*},\phi_{<}^{*}\right]\right].
\end{align*}
Thus the change in the action after integrating out the high-momentum
modes is given by, using that the trace is over an infinitesimal momentum
interval $\Lambda<k<\Lambda+\mathrm{d}\Lambda$,
\begin{equation}
\mathrm{d}S=\frac{\hbar}{2}\mathrm{Tr}\delta_{\Lambda}\ln\left[-G_{0}^{-1}+\frac{1}{\hbar}\frac{\delta S_{\mathrm{int}}}{\delta\mathbf{\Phi}\delta\mathbf{\Phi}^{\dagger}}\right]\mathrm{d}\Lambda.
\end{equation}

\section{Ultra-violet subtractions\label{sec:UVsubtr}}

To calculate the condensate density and total density an ultra-violet
subtraction is necessary, see Eq.~(\ref{eq:cond_dens}) and Eq.~(\ref{eq:Total_density-1}).
This subtraction is a consequence of the renormalization of the bare
coupling to the two-body $T$ matrix $T^{\mathrm{2B}}(-2mc^{2})$
\citep{Stoof2009Ultracold}. The phase-fluctuation and non-phase-fluctuation
propagator in real space are written as
\begin{align}
\left\langle \theta(\mathbf{k},\omega_{n})\theta(\mathbf{k},\omega_{n})\right\rangle  & -\frac{\frac{1}{n}mc^{2}}{(\hbar\omega_{n})^{2}+\left(\epsilon_{\mathbf{k}}+mc^{2}\right)^{2}},\label{eq:UVsubtrPhasefluc}\\
\left\langle \phi'(\mathbf{k},\omega_{n})\phi'^{*}(\mathbf{k},\omega_{n})\right\rangle  & +\frac{mc^{2}}{(\hbar\omega_{n})^{2}+\left(\epsilon_{\mathbf{k}}+mc^{2}\right)^{2}},\nonumber
\end{align}
where $mc^{2}=n_{0}g$ and the propagators are defined in Eqs.~(\ref{eq:Phase_fluc_prop-1})
and (\ref{eq:High_En_fluc-1}). This implies that the equal-time correlation
function $\left\langle \theta(\mathbf{x},0)\theta(\mathbf{0},0)\right\rangle $
with ultra-violet subtraction at zero temperature is given by
\[
\frac{1}{V}\left(\frac{1}{n}mc^{2}\right)\sum_{\mathbf{k}}\left(\frac{1}{2\hbar\omega_{\mathbf{k}}}-\frac{1}{2(\epsilon_{\mathbf{k}}+mc^{2})}\right)\cos(\mathbf{k}\cdot\mathbf{x}).
\]
In the long-range limit ($|\mathbf{x}|\rightarrow\infty$) we have
that this expression vanishes, which is used to define the condensate
density in Eq.~(\ref{eq:cond_dens}). Whereas the equal-time correlation
function of the non-phase fluctuations $\left\langle \phi'(\mathbf{x},0)\phi'^{*}(\mathbf{0},0)\right\rangle $
with the ultra-violet subtraction at zero-temperature is
\begin{equation}
\frac{1}{V}\sum_{\mathbf{k}}\left[\frac{\epsilon_{\mathbf{k}}-\hbar\omega_{\mathbf{k}}}{2\hbar\omega_{\mathbf{k}}}+\frac{mc^{2}}{2(\epsilon_{\mathbf{k}}+mc^{2})}\right]\cos(\mathbf{k}\cdot\mathbf{x}).\label{eq:non_phas_fluc_UVsubtr}
\nonumber
\end{equation}
The contribution to the total density due to non-phase-fluctuations
follows from evaluating this expression at equal position $n'=\left\langle \phi'(\mathbf{0},0)\phi'^{*}(\mathbf{0},0)\right\rangle $.

\bibliographystyle{apsrev4-1}
\bibliography{bose_gas}

\end{document}